\documentclass[prc,twocolumn,showpacs]{revtex4}
\usepackage{graphicx}
\usepackage{bm}

\begin{document}
\title{Landau parameters of nuclear matter in the 
           spin and spin-isospin channels}
\author{W. Zuo$^{1}$, Caiwan Shen$^{2,3,4}$ and U. Lombardo$^{2,5}$}
\affiliation{
   $^{1}$ Institute of Modern Physics, Chinese Academy 
          of Sciences, Lanzhou, China \\
   $^{2}$ INFN-LNS, Via S. Sofia 44, I-95123 Catania, Italy \\
   $^{3}$ China Institute of Atomic Energy, P.O.Box 275(18), 
          Beijing 102413, China \\
   $^{4}$ Center of Theoretical Nuclear Physics, National Laboratory of 
          Lanzhou Heavy Ion Accelerator, Lanzhou, China \\
   $^{5}$ Dipartimento di Fisica, Via S. Sofia 64, I-95123 Catania, Italy
   }

\begin{abstract}
The equation of state of spin and isospin polarized nuclear matter is 
determined in the framework of the Brueckner theory including three-body 
forces. The Landau parameters in the spin and spin-isospin sectors are 
derived as a function of the baryonic density. The results are compared 
with the Gamow-Teller collective modes. The relevance of $G_0$ and $G_0'$ 
for neutron stars is shortly discussed, including the magnetic 
susceptibility and the neutron star cooling.
\end{abstract} 


\maketitle

The equation of state (EOS) of nuclear matter is a major issue of the 
quantum-mechanical many-body problem. The reliability of any theoretical
prediction has been measured on its capability of reproducing the 
empirical saturation energy and density, once the convergence of the 
theory has been firmly established. This is the case of the 
Brueckner-Bethe-Goldstone theory: the hole-line expansion has been 
proved in fact to be rapidly converging \cite{SONG}, and the 
Brueckner-Hartree-Fock (BHF) approximation implemented by  
three-body forces (3BF) can account for the empirical saturation 
point \cite{UMB}. Besides saturation density and energy additional 
constraints have been the compressibility extracted from the monopole 
energy and the symmetry energy from the binding energy of $N\neq Z$ nuclei. 
In the spin-isospin channel a further constraint has been provided by 
the Gamow-Teller (GT) giant resonances (see \cite{OSTE} for a review). 
Recently the Landau parameter $G_0'$ has been extracted from the 
experimental data with a very small uncertainty \cite{SAK}. So it 
represents a very robust constraint for the EOS of nuclear matter. 
The theoretical prediction of $G_0'$ demands for extending the 
calculation of the EOS of nuclear matter to spin-polarized neutrons 
and protons. Such calculations have been stimulated by the search of 
a spontaneous transition to a ferromagnetic state to explain the 
strong magnetic fields observed in neutron stars \cite{VIDA,FRASC,BOMB}, 
but it can have important implication in the physics of atomic nuclei. 
 
We extended the calculation of the EOS to polarized nuclear matter 
in the framework of the BHF approximation with the continuous choice for the 
auxiliary potential. As 2BF we took the Argonne $AV_{18}$  
\cite{AV18} and as 3BF the one from Ref.\cite{TBF}. Starting from unpolarized 
nuclear matter at a given density $\rho$ (we only consider 
symmetric nuclear matter), we ran different 
polarization states of neutrons and protons
\begin{eqnarray}
\delta_n =\frac{N_\uparrow - N_\downarrow}{N} \quad \delta_p 
=\frac{Z_\uparrow - Z_\downarrow}{Z}\quad \rho=\frac{N+Z}{V}.
\end{eqnarray} 
The results are reported in Fig.~1 for two typical densities of nuclear
matter with 2BF (left side) and 2BF plus 3BF (right side).
Since the EOS of isospin-polarized  nuclear matter fulfills a quadratic 
law as a function of isospin-symmetry parameter \cite{ISO,ISOS}, and the 
same is true for the EOS of spin-polarized nuclear matter versus the 
spin-symmetry parameter \cite{FRASC,BOMB}, 
we fit the EOS (data plotted as symbols in Fig.1) by the least square 
method in the mixed case according to a quadratic law 
\begin{equation}
E_A(\rho,\delta_n,\delta_p)-E_A(\rho,0,0) = \sum_{\tau\tau'} \Lambda_{\tau,\tau'}(\rho)
\delta_{\tau}\delta_{\tau'}, 
\end{equation} 
where $\tau=n,p$ is the isospin quantum number. 
\begin{figure}
\centering{\includegraphics[width=85mm]{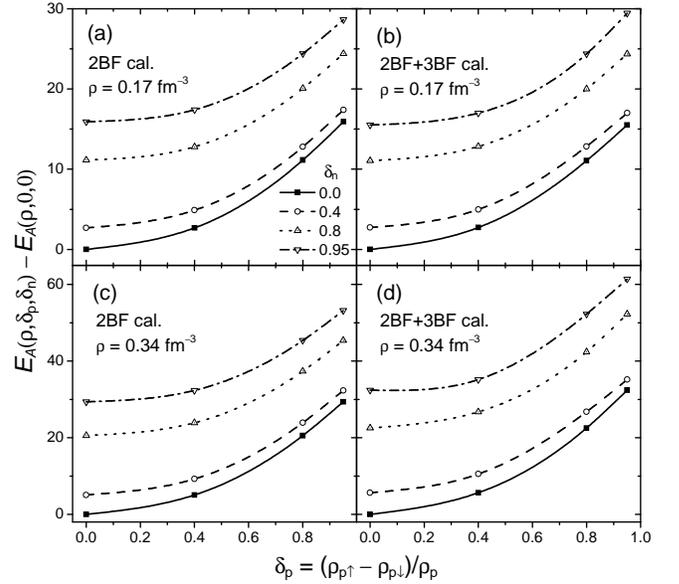}}
\caption{EOS of spin-polarized nuclear matter. The symbols are from 
microscopic calculations, and the lines are drawn only to guide the eye.}
\end{figure}
In symmetric nuclear matter (SNM) the coefficients $\Lambda_{\tau\tau'}$ 
are related to the zero-order Landau parameters
\begin{eqnarray}
G_0 = G_{nn}^{0} + G_{np}^{0} = \frac{4N(0)}{\rho}(\Lambda_{nn}+\Lambda_{np}) - 1, \\
G'_0 = G_{nn}^{0} - G_{np}^{0} = \frac{4N(0)}{\rho}(\Lambda_{nn}-\Lambda_{np}) - 1, 
\end{eqnarray}
where $N(0)$ is the level density at the Fermi surface. 
In Fig.~2 the two Landau parameters are plotted as a function of the 
density based on Eqs.(3) and (4).
\begin{figure}
\centering{\includegraphics[width=8cm,angle=0]{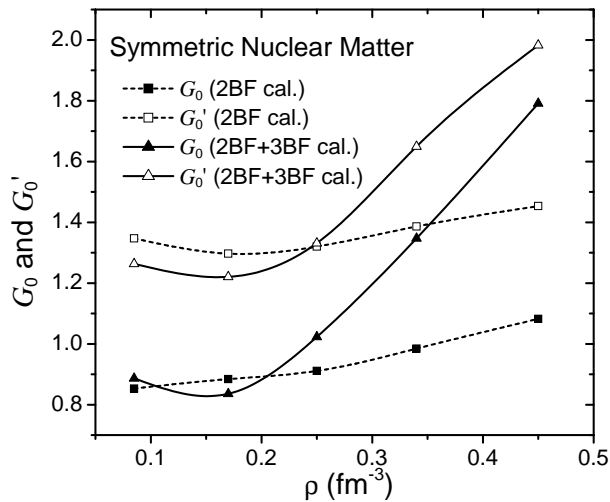}}
\caption{Landau parameters of SNM in the spin and spin-isospin channel.
         The square(tri-angle) symbols are from 2BF (2BF+3BF) calculations 
         and the solid(open) ones are for $G_0$ ($G_0'$). }
\end{figure}

The Landau parameter $G'_0$ is the strength of the spin-isospin component
$V_{\sigma\tau} = G'_0(\sigma_1\cdot\sigma_2)(\tau_1\cdot\tau_2)$ of the 
residual interaction\cite{MIG}, which governs the Gamow-Teller (GT) giant 
resonance in nuclei \cite{OSTE}.  
Its value at the saturation point  has been determined with high precision 
from the experimental excitation energy of the GT resonance 
on $^{90}$Ni \cite{SAK}. The value reported is $1.182 < G'_0 <1.188$ 
(we have multiply by a factor two according 
to the definition we adopted). More recent fit on $^{112}$Sn and $^{208} $Pb
within a RPA calculation with Skyrme forces confirm such a prediction\cite{BEN}. 
Our nuclear-matter prediction of $G_0'$  at the saturation density, 
which is about 1.22 including 3BF is in a pretty good agreement with the   
previous values. The value without 3BF of about 1.30 is in less 
agreement. Since the behaviour of $G'_0$ around
the saturation point is very flat, there is no room  for large uncertainties 
in the comparison. Such an agreement provides a further support  to the 
important role played by the microscopic 3BF as to 
reproducing all saturation properties 
of nuclear matter. Other predictions of $G_0'$ including 3BF are from 
phenomenological Skyrme forces, which unfortunately are scattered in
wide range of values  lower than the experimental one \cite{BEN}.

So far experimental information on $G_0$ is not enough since spin 
resonances have only been observed with too small strength compared 
to other collective modes\cite{OSTE}.  
  
The prediction of $G_0$ and $G'_0$ for densities other than the 
nuclear density, which is reported in Fig. 2, is of great 
interest in the study of neutron
stars. In connection with the strong magnetic fields 
observed in neutron stars some
authors\cite{FANT,VIDA,FRASC,BOMB} studied the magnetic susceptibility 
$\chi$ in neutron matter and found that $G_0$ reduces the $\chi$ of 
degenerate neutron gas. This reduction is amplified at high density 
when including 3BF either in Brueckner calculations\cite{FRASC} and in 
Montecarlo many-body simulations \cite{FANT}.

Spin and spin-isospin excitations of nuclear matter are  coupled to   
the weak interaction governing the neutrino emission of URCA processes 
as well as the neutrino transport in neutron stars. The high-density 
increase  of $G_0$ and $G_0'$, driven by the 3BF, is expected to have 
important implications for the neutron star cooling  for the
sizeable enhancement induced  by the nuclear 
correlations on the neutrino mean free path \cite{GIAI}.

In conclusion, in this note we reported on a BHF 
calculation of the Landau parameters
$G_0$ and $G_0'$ as a function of baryonic density. 
The main scope was to point out the large effect of 3BF, especially at high densities.
At the 2BF level, there is a wide disagreement with previous Brueckner calculations 
(see Ref.\cite{BALDO} and Refs. therein quoted) that has 
not yet clearly explained, since one can hardly
control and compare the different approximations. 
On the other hand, our prediction for $G_0'$ has been
found to be in very good agreement with the experimental value 
extracted from GT resonance when 3BF is included. The 
relevance of the spin Landau parameter for neutron stars has been also 
discussed in connection with magnetic susceptibility and neutrino
mean free path.

This work is supported in part by the Chinese Academy of Science within the
{\it One Hundred Person Project} and the NNSF of China under the contract No.s 
10075078, 19935030 and 10047001.

\end{document}